\documentclass[aps,pra,twocolumn,showpacs,preprintnumbers,longbibliography]{revtex4-1}

\usepackage{psfrag,graphicx}
\usepackage{dcolumn}
\usepackage{amsmath,amssymb}
\usepackage{bm}
\usepackage{amsfonts,amssymb,amsmath}        
\usepackage{epstopdf}
\usepackage{enumerate}
\usepackage{float}
\usepackage{color}

\newcommand{\be}{\begin{equation}}
\newcommand{\ee}{\end{equation}}
\newcommand{\bq}{\begin{eqnarray}}
\newcommand{\eq}{\end{eqnarray}}

\newcommand{\ket}[1]{\left | \, #1 \right\rangle}
\newcommand{\bra}[1]{\left \langle #1 \, \right |}
\begin{document}

\title{Error correction for non-Abelian topological quantum computation}
\author{James R.~Wootton$^1$, Jan Burri$^1$, Sofyan Iblisdir$^2$, Daniel Loss$^1$}
\affiliation{$^1$Department of Physics, University of Basel, Klingelbergstrasse 82, CH-4056 Basel, Switzerland\\
$^2$University of Barcelona, Dpt. Estructura i Constituents de la Materia, 647 Av. Diagonal, 08028 Barcelona, Spain}

\date{\today}

\begin{abstract}

The possibility of quantum computation using non-Abelian anyons has been considered for over a decade. However the question of how to obtain and process information about what errors have occurred in order to negate their effects has not yet been considered. This is in stark contrast with quantum computation proposals for Abelian anyons, for which decoding algorithms have been tailor-made for many topological error-correcting codes and error models. Here we address this issue by considering the properties of non-Abelian error correction in general. We also choose a specific anyon model and error model to probe the problem in more detail. The anyon model is the charge submodel of $D(S_3)$. This shares many properties with important models such as the Fibonacci anyons, making our method applicable in general. The error model is a straightforward generalization of those used in the case of Abelian anyons for initial benchmarking of error correction methods. It is found that error correction is possible under a threshold value of $7 \%$ for the total probability of an error on each physical spin. This is remarkably comparable with the thresholds for Abelian models.

\end{abstract}
\maketitle

\section{Introduction}

Topological quantum computation is the proposed use of anyonic quasiparticles to non-locally store quantum information and process it in a way that is immune to the effects of small perturbations \cite{preskill,brennen,pachos_book,wootton_rev}. As such, it is often said to be `inherently fault-tolerant' and `topologically protected'. However, it is important to note that this topological protection does not replace the need for active error correction. In fact, the topological protection only arises when error correction is employed. For anyonic systems, this correction involves continual measurement of anyon occupations to determine when and where errors cause unwanted anyons to be created. These results must be classically processed in order to determine how these anyons can be removed without disturbing the computation. We refer to the combination of the measurements and processing as the decoding algorithm.

For schemes based on non-Abelian anyons \cite{double,mochon,magic,fibonacci,wootton_ds3}, the requirement for active error correction is often ignored. The presence of an energy gap that suppress the creation of unwanted anyons may seem to replace the need for error correction. However, an $O(1)$ gap will only suppress anyon creation to a limited extent. Once the computation becomes sufficiently large, in terms of either the size of the quantum computer or its runtime, the presence of unwanted anyons becomes almost certain. The computation will then be unable to proceed correctly. The lack of error correction then means the lack of scalability, which is a fundamental requirement for a true quantum computer.

This is in stark contrast to computation schemes based on Abelian anyon models \cite{raussendorf,fowler_rev,wootton_abelian,wootton_thesis}. For these the fact that the topological protection arises only when using active error correction has long been understood. As such the required methods to achieve error correction have been intensively studied and many efficient and effective decoding algorithms have been produced \cite{dennis,fowler,renorm,wootton,broom,hutter,renorm_d,fowler_new,wootton13}.

Research into the corresponding methods required for non-Abelian error correction is long overdue. Even a basic understanding of the decoding problem is yet to be developed. This is a serious issue for proposals based on non-Abelian models, since without proper proof that the required error correction is possible it is hard to argue that these proposals could truly realize fault-tolerant quantum computation.

In this paper we seek to address this important issue. We consider the problem of error correcting non-Abelian anyon models in general, determining its similarities and fundamental differences to the Abelian case. We therefore lay out the framework for future study of non-Abelian error correction towards the goal of a full demonstration that fault-tolerance is possible. We also choose a specific anyon model and error model to probe the problem in more detail. The anyon model chosen is the charge submodel of $D(S_3)$ or $\Phi-\Lambda$ model \cite{double,brennen,wootton_ds3}. This shares many important properties with Fibonacci anyons, and all other models that are known to be universal for quantum computation. It is also known to be universal itself when supplemented with non-topological operations. The error model is a straightforward generalization of those used in the case of Abelian anyons for initial benchmarking of error correction methods. It is shown that error correction is indeed possible in this case as long as the total probability of error on each spin is below $7 \%$. Remarkably, this threshold value is almost identical to that obtained for Abelian anyons using the same method \cite{wootton13}.

\section{Abelian and non-Abelian anyons}

Anyons belong to two classes: Abelian and non-Abelian. The class of Abelian anyons is defined to hold all particle types for which fusion with any other type yields a single definite result. Non-Abelian anyons are such that fusion with at least one other particle type yields a multiplicity of results. Any anyon model that contains at least one type of non-Abelian anyon is called a non-Abelian anyon model.

In general, most non-Abelian anyon models have fusion rules in which the fusion of two non-Abelian anyons can yield further non-Abelian anyons. This fact is particularly relevant from the perspective of quantum computation, since all known models that can achieve universality using topological operations alone have this kind of behaviour. This includes the Fibonacci anyon model \cite{fibonacci} and non-Abelian quantum double models \cite{double,mochon}. However, this kind of behaviour is completely opposed to the simplicity found in Abelian models. This will likely mean that the decoding of non-Abelian anyons is much more complex than the Abelian case in general, and may even mean that successful decoding is not possible. If true, this will have dire effects for many proposals for topological quantum computation with non-Abelian anyons.

It is therefore important to begin the investigation into the decoding of anyon models with this general form of behaviour. Here we investigate one of the simplest such models, and the only known one to be well described by a classical simulation. This of course means that it is not universal for quantum computation by topological operations alone. However it can be made universal with the addition of straightforward non-topological elements \cite{wootton_ds3}. A simple error model is considered, and numerical evidence is presented to show that it is indeed possible to successfully decode the model below a threshold noise rate.

It is worth noting that there are some non-Abelian anyon models for which the fusion of two non-Abelian anyons always yields an Abelian anyon. The most prominent example is the Ising anyon model \cite{honeycomb}, which is not universal by topological operations alone. The simple structure of the fusion rules makes the problem of decoding this model quite different to the general case \cite{talk}.

\section{$\Phi-\Lambda$ model} \label{sec:phi-lambda}

An important class of non-Abelian anyon models that may be realized on spin lattice systems are those of the quantum double construction \cite{double,brennen}. Each of these are based on a non-Abelian group, $G$, and is then referred to as the $D(G)$ anyon model (Abelian anyon models can be similarly constructed from Abelian groups). In all quantum double models, all types of anyon fall into three classes: charge, flux and dyon. Fusion of charge anyons with each other will only ever yield charge anyons or the vacuum. They therefore form their own fully complete and consistent anyon model: the charge sub-model of $D(G)$.

The simplest non-Abelian group is the permutation group of three objects, $S_3$. The simplest non-Abelian quantum double anyon model is therefore $D(S_3)$. Even so, it is a highly complex anyon model with many different anyon types. The ability to further simplify by considering only the charge sub-model, which has only two non-trivial types of anyon, is therefore highly advantageous. It is this that we will consider here. We refer to the charge sub-model of $D(S_3)$ as the $\Phi-\Lambda$ model, in reference to the common labels for its two anyon types $\Phi$ and $\Lambda$.

To define an anyon model, one starts by listing all possible possible particle types. Since this list is complete, all composite particles must also behave according to one of these types. The rules by which this occurs are known as the fusion rules. When the composite of particles of type $a$ and $b$ can behave either as type $c$ or $d$, the rule is written $a \times b = c + d$. The orders of factors and terms in these rules holds no physical relevance. The composite of any particle type with the vacuum, which is usually denoted $1$, always yields the original type. Such rules are therefore usually left unwritten. The fusion rules can also be read backwards as splitting rules, and so $a \times b = c + d$ implies that both a particle of type $c$ or a particle of type $d$ can split into a pair of particles consisting of one $a$ and one $b$.

For the $\Phi-\Lambda$ model, the three possible particle types are the vacuum, $1$, an Abelian anyon, $\Lambda$, and a non-Abelian anyon, $\Phi$. Their non-trivial fusion rules are,
\be \label{rules}
\Lambda \times \Lambda = 1, \,\, \Phi \times \Lambda = \Phi, \,\, \Phi \times \Phi = 1 + \Lambda + \Phi.
\ee
Note that both $\Lambda$ and $\Phi$ are their own antiparticles, since fusion of either type with itself can yield $1$. However, the composite of two $\Phi$'s sometimes results in a $\Lambda$ or even another $\Phi$ rather than annihilation to the vacuum.

\begin{figure}
\begin{center}
\includegraphics[width=7.5cm]{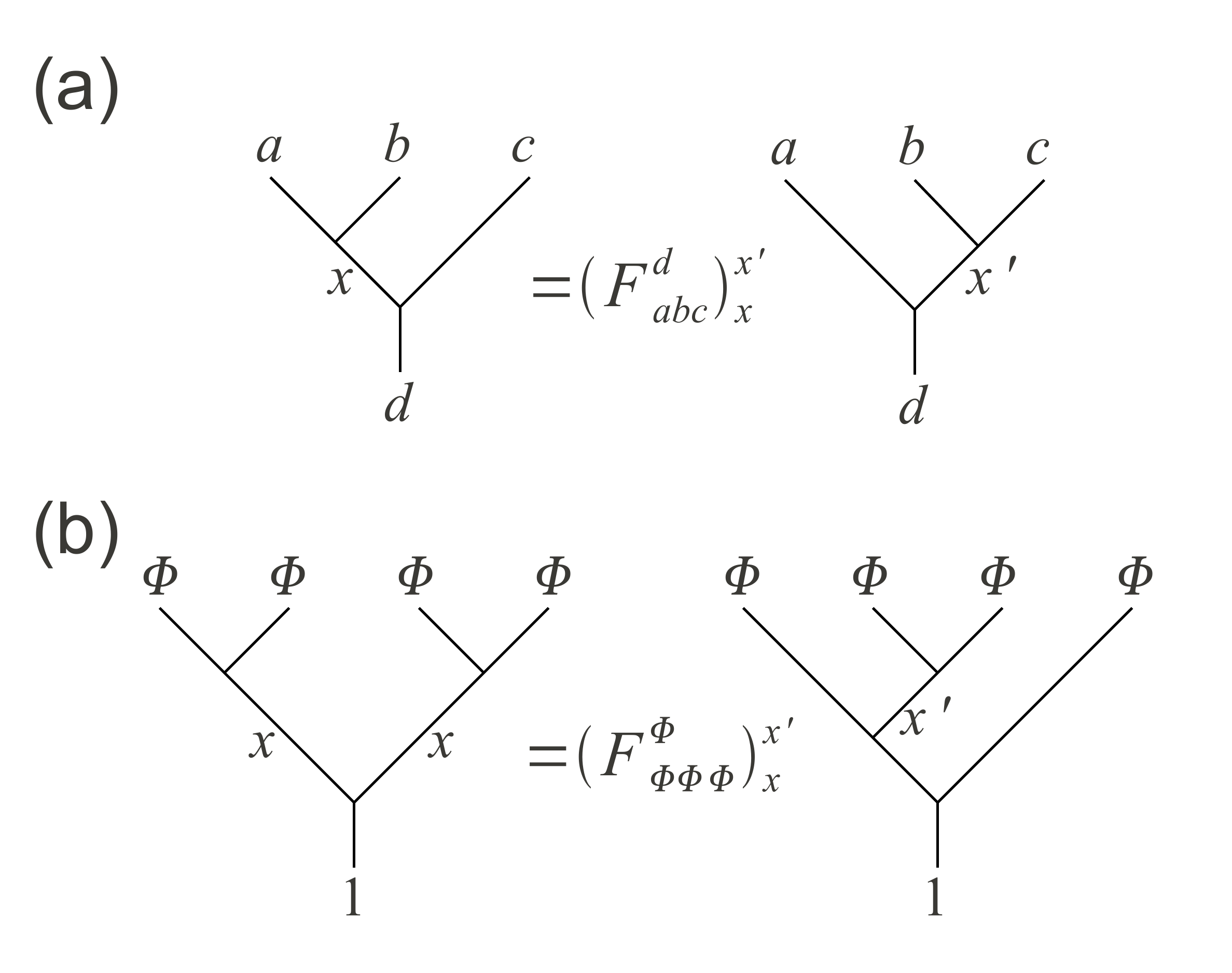}
\caption{\label{F&R}
(a) The two fusion bases for three anyons $a$, $b$ and $c$ created from an anyon $d$. The definition of the $F$-matrix that relates these is shown. (b) The two fusion bases for four $\Phi$ anyons created from the vacuum. One is depicted to the left of the equation, whose intermediate step is labelled by $x$. The other, to the right, has intermediate step $x'$. These are related by the same $F$-matrix as three $\Phi$ anyons created from one $\Phi$ as shown.}
\end{center}
\end{figure}

There are many distinguishable processes by which $n$ particles can be created from a single particle (such as the vacuum). The resulting states belong to the so-called fusion space of the $n$-anyons. Some of the processes differ only in the anyon types used for intermediate steps. The resulting states are referred to as different fusion states within the same fusion basis. When a completely different process is used, the resulting fusion states belong to a different fusion basis. See Fig. \ref{F&R} for an illustration.

There are also many ways by which $n$ particles may be fused back the original particle type. These are the reverse of the processes used to create the anyons and correspond to a measurement of the particles. The process of fusion determines the fusion basis that is measured in, and the identity of the intermediate particles determine the specific outcomes that result.

The relationships between all the many possible fusion bases for all the many possible collections of anyons can be determined by a few simple unitary transformations: the $F$-matrices. These are the basic building blocks of basis transformation from which all others can be derived. Their definition is shown in Fig. \ref{F&R}. For the $\Phi-\Lambda$ model there is only one non-trivial $F$ matrix. This describes the transformation when three $\Phi$'s have been created from a single $\Phi$, and takes the form,
\be \label{ChargeF}
F_{\Phi\Phi\Phi}^{\Phi} = \frac{1}{2}\left(
\begin{array}{rrr}
1 & 1 & -\sqrt{2} \\
1 & 1 & \sqrt{2} \\
-\sqrt{2} & \sqrt{2} & 0
\end{array} \right),
\ee
in the basis $1$, $\Lambda$, $\Phi$ for the intermediate result \cite{nonlocal}.

The effect of exchanging two adjacent anyons $a$ and $b$ is twofold. Firstly there is the permutation of the particle types, and secondly a phase will be acquired. The latter depends on what would be the fusion outcome if $a$ and $b$ were fused. In the case that $a$ and $b$ fuse to $c$, the phase is written $R_{a b}^c$. For the $\Phi$'s, these phases are $R_{\Phi \Phi}^1 = R_{\Phi \Phi}^{\Phi} = 1$ and $R_{\Phi \Phi}^{\Lambda} = -1$.

For most non-Abelian anyon models, the effect of exchanges not only represents the permutation of the particles, but also leads to additional rotations within the fusion space. However, this is not the case for this model, where the exchanges have no effect beyond permutation. Even so they do affect the fusion behaviour. In fact, these effects mean that the braiding is a non-Abelian representation of the permutation group. As such, the $\Phi-\Lambda$ model is rightly called a non-Abelian anyon model. A more detailed demonstration of the fusion and braiding properties of this model can be found in Appendix \ref{app:phi-lambda}.

To show that the decoding methods considered for this model also have relevance to other non-Abelian models, let us consider the Fibonacci anyon model. This model has a single non-trivial anyon type, $\tau$, and a single non-trivial fusion rule $\tau \times \tau = 1 + \tau$. Using the correspondence $\tau \rightarrow \Phi$ we can see that this is the same as the $\Phi \times \Phi$ rule in the $\Phi-\Lambda$ model, except for the addition of the Abelian $\Lambda$ anyon in the latter. Therefore, if one decodes the $\Phi-\Lambda$ model by first considering only the $\Phi$'s until none remain, and only then considering the $\Lambda$'s, the decoding algorithm is also directly applicable to Fibonacci anyons. This point is expanded upon in Appendix \ref{app:fibonacci}.

\section{Classical simulation of the non-Abelian anyons}\label{sec:classical}

\begin{figure}[h]
\begin{center}
\includegraphics[width=6cm]{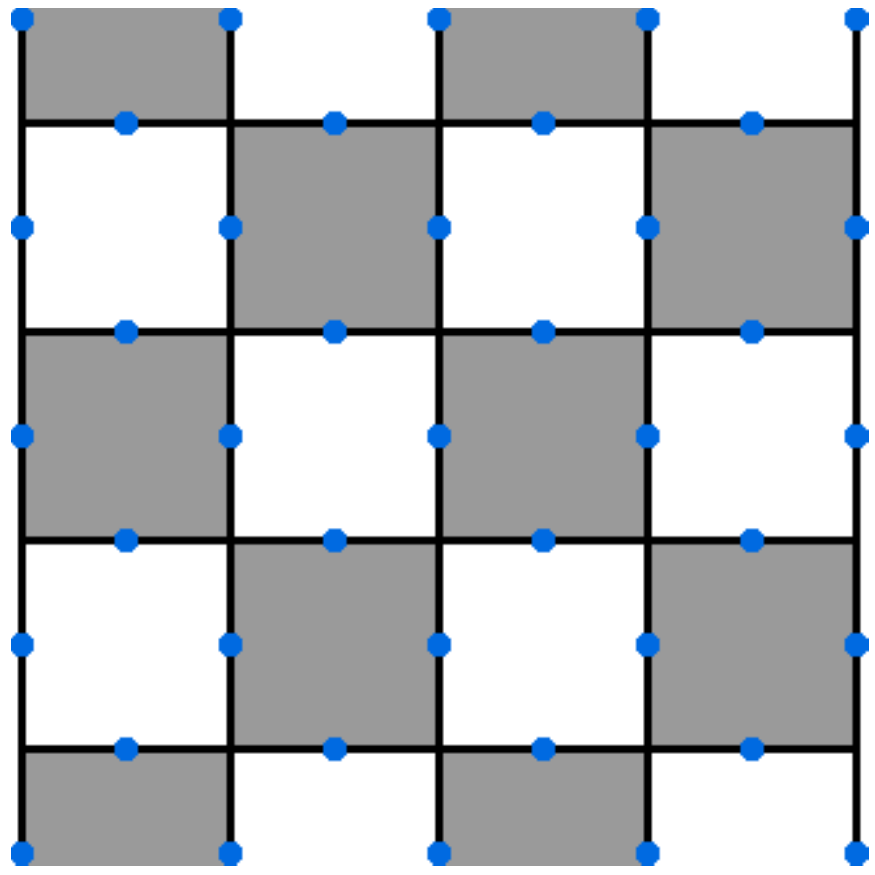}
\caption{\label{lattice}
Lattice used to simulate the $\Phi-\Lambda$ model. Six-level spins, shown by blue dots, are placed on each edge. Plaquettes are bicoloured white and grey in chessboard fashion. The size of the lattice is measured by $L$, the number of spins along each edge. $L=5$ in the lattice shown.}
\end{center}
\end{figure}

The standard means by which the $\Phi-\Lambda$ model can be realized on a spin lattice is by using the quantum double construction of \cite{double,brennen} for the group $S_3$. This results in a complex and highly entangled spin model. However, here we define a classical model based on the $Z_6$ quantum double model (itself a good error correcting code) that allows an efficient simulation of these anyons. This classical model used is defined on the spin lattice shown in Fig. \ref{lattice}, where a six-level spin is placed on each edge. The state of the $j$th spin is denoted $\ket{s_j}$ and will always be one of the basis states $\{ \ket{0}, \ket{1} \, \ldots, \ket{5}\}$. We will consider the following unitary operator on these spins,
\be
R^g_j = \sum_{s_j} \ket{s_j + g \mod 6}\bra{s_j}.
\ee
This rotates a spin state $s_j$ to $s_j+g \mod 6$. Clearly $R^{6-g} = (R^g)^{-1}$.

There are two types of plaquette, bicolored white and grey in chessboard fashion. For white plaquettes we associate a value $\sigma = +1$ and for grey we have $\sigma = -1$. For each plaquette, $p$, we define a variable $b_p$ as follows,
\be
b_p = \sigma_p \sum_{j \in p} s_j  \mod \, 6.
\ee
Here the summation runs over all spins, $j$, around the plaquette $p$. For the majority of plaquettes there are four such spins, but the plaquettes on the top and bottom edges have only three. When calculating the sum we use the convention that $-x \mod 6 = 6n-x \mod 6$, where $n$ is an integer that satisfies $6n \geq x$. Similar variables $b_l$ and $b_r$ are also defined for the left and right edges. These are calculated using a sum over all $L$ spins on the extreme left and right edges of the code, respectively. The term contributed by the spin $j$ is $s_j$ if $j$ is part of a grey plaquette and $-s_j$ otherwise. Equivalent but independent variables can be defined for the vertices and top and bottom edges. However, since these are equivalent, it is sufficient to consider only the plaquettes.

The value of $b_p$ is used to define the anyon occupancy of the plaquette $p$ (or edge $l$ or $r$). If $b_p=0$ the plaquette is said to hold only vacuum. If $b_p=3$ the plaquette is said to hold a $\Lambda$. Otherwise ($b_p \in \{1,2,4,5\}$) the plaquette is said to hold a $\Phi$. As such we will use the notation $\phi = \{1,2,4,5\}$ from now on.

Consider the state for which all plaquettes hold only vacuum. When defined properly, this total vacuum state should be an equally weighted mixture of all topologically equivalent states with $b_p=0$ $\forall p$. This is in order to hide the `strings' of anyons once they are created.

Any spin $j$ will be part of two plaquettes, one white and one grey. The application of $R^3_j$ to the vacuum state will change the value of $b_p$ on both these two plaquettes to $3$. The effect of the rotation is therefore to create a pair of $\Lambda$ anyons. Similarly, application of $R^g_j$ for $g \in \phi$ will change $b_p$ on the white plaquette to $g$ and that on the grey to $6-g$. This creates a pair of $\Phi$ anyons. Further applications of these operations on spins can be used to move the anyons around. Note that when a $\Phi$ on plaquette $p$ with $b_p=\beta$ is moved to an initially empty plaquette $p'$, this will result in $b_p=0$ (reflecting the fact that $p$ is now empty) and $b_{p'}=\beta$. As such, the value $\beta$ is an internal state of the anyon that is carried with it.

If two anyons are moved to the same plaquette, the resulting occupation of that plaquette will be the result of their fusion. When an anyon $i$ with internal state $\beta_i$ meets an anyon $j$ with state $\beta_j$, the result is an anyon $k$ with $\beta_k = \beta_i + \beta_j \mod 6$. From this it is clear to see that two $\Lambda$s will annihilate, a $\Lambda$ fused with a $\Phi$ always results in a $\Phi$, and two $\Phi$s will either annihilate, form a $\Lambda$, or form another $\Phi$. This behaviour exactly reproduces the fusion rules of Eq. (\ref{rules}).

For a good simulation of the $\Phi-\Lambda$ model, it is important to reproduce the correct statistics as well as the correct fusion rules. In order to do this, randomness must be incorporated into the way in which a single $1$, $\Lambda$ or $\Phi$ splits into a pair of $\Phi$'s. To do this, the state $\rho(\Phi,\Phi|x)$ of two $\Phi$ anyons that were split out of an anyon type $x$ is defined to be,
\bq\nonumber \label{create_probs}
\rho(\Phi,\Phi|1) &=& \frac{1}{4} \sum_{j \in \phi} \ket{ \beta_j , \beta_{-j} } \bra{ \beta_j , \beta_{-j} }, \\ 
\rho(\Phi,\Phi|\Lambda) &=& \frac{1}{4} \sum_{j \in \phi} \ket{ \beta_j , \beta_{3-j} } \bra{ \beta_j , \beta_{3-j} }, \\ \nonumber
\rho(\Phi,\Phi|\Phi) &=& \frac{1}{8} \sum_{i \in \{1,2\} } \, \sum_{j,k \in \{0,3\}} \ket{ \beta_{i+j} , \beta_{i+k} } \bra{ \beta_{i+j} , \beta_{i+k} }.
\eq
In other words, to split two $\Phi$'s out of a $1$ or $\Lambda$ anyon in a plaquette $p$, the rotation $R^j_i$ should be applied to one of the spins around $p$ with the value of $j \in \phi$ chosen randomly. If it is a $\Phi$ in $p$, corresponding to a value $\beta_k$, the value of $j$ should be chosen randomly from the values $\{1,4\}$ for $k\in \{2,5\}$ or vice-versa.

Note that the randomness in each $\Phi$ anyon means that they can only be manipulated using controlled operations. The splitting of one $\Phi$ to two involves such an operation, since the rotation $R^j_i$ used depends on the internal state of the $\Phi$. Clearly the same is true for the operation required to move a $\Phi$. It is important that the internal state is not recorded during such operations in order to maintain the randomness.

Demonstrations of how this classical model reproduces the behaviour of the $\Phi-\Lambda$ model can be found in the Appendices. A plausibility argument, using only concepts introduced in here and in Section \ref{sec:phi-lambda}, can be found in Appendix \ref{app:phi-lambda}. A more rigorous proof using the structure of the $D(S_3)$ lattice model and the error model of the following section can be found in Appendix \ref{app:DS3}.

\section{Error Model} \label{sec:error}

For the purposes of this study we consider that, as in the standard $D(Z_2)$ planar code \cite{dennis}, information is stored using states for which all plaquettes are empty of anyons, and different anyon occupancies of the edges are used for the different logical states. The effect of noise is then to apply operations randomly to the spins, creating $\Phi$ and $\Lambda$ anyons on plaquettes and changing the anyon occupancy of the edges. To undo the effects of the errors, the resulting anyon configuration on the plaquettes must be measured and the result processed. Further rounds of measurement and classical processing may also be required. The desired end result is to determine how to annihilate the anyons in a way that is topologically equivalent to the means by which they were created. In this case, topological equivalence means that the final edge occupancies are the same as their initial values. The stored information will then be preserved, without the process ever needing to measure and disturb it. More details on the edges for a specific lattice model that realizes the $\Phi-\Lambda$ anyons can be found in \cite{boundary}.

The success of the decoding procedure is measured by the so-called logical error rate, $P$. This is the probability that the decoding procedure does not annihilate the anyons in a manner that is topologically equivalent to their creation. A good decoder should achieve a logical error rate of $P = O( \exp{[-\alpha(p) L^{\beta}]})$ for positive $\beta = O(1)$. Here $p$ denotes the strength of the noise that creates the anyons. The exact interpretation of $p$ depends on the error model used. The quantity $\alpha(p) \geq 0$ governs the decay of the logical error rate. If successful decoding is possible, there will be a finite threshold value $p=p_c$ such that $\alpha(p) > 0$ $\forall p<p_c$.

We will consider errors that act independently on each spin and apply `flip' operations of the form $R^j_i$ (we can consider an equivalent but independent anyon model defined using the vertices of the lattice to equivalently and independently deal with phase errors). In order to be consistent with Eq. (\ref{create_probs}) and maintain the proper simulation of the anyon, we must ensure that the probability for the spin flips $R^j_i$ are equal for $j \in \phi$. We will use $p_{\Phi}$ to denote the probability that one of these flips is applied, and so each will be applied with probability $p_{\Phi}/4$. We will use $p_{\Lambda}$ to denote the probability that the flip $R^3_i$ is applied to each spin. The probability $1-p_{\Phi}-p_{\Lambda}$ is therefore the probability that a spin suffers no flip. In this study we will consider the case of $p=p_{\Phi}=p_{\Lambda}$. The total probability of an error on each spin is then $2p$.

\section{The decoding algorithm}

The nature of syndrome extraction in the non-Abelian decoding problem means that fusion measurements must be made in order to provide sufficient information to decode. The most subtle way to do this, extracting as much information as possible, is to fuse pairs of anyons (as opposed to fusing clusters). However, these fusions cannot be chosen arbitrarily (as explained later in Section \ref{sec:nature}) but must instead be chosen carefully according to the likely error configurations. We consider one means to do this, though others are also possible.

The decoding algorithm used is that studied in \cite{wootton13}. It is applied in this case as follows,

\begin{enumerate}

\item Loop through all plaquettes to find $\Phi$ anyons. By convention, loop from left to right and top to bottom.

\item For each anyon, search through all plaquettes at a Manhattan distance of $k$. Initially $k=1$.

\item If another $\Phi$ anyon (or the edge) is found at this distance, pair them. Move one to fuse with the other by performing the required controlled $R^j_i$ rotations on the connecting spins. The direction of the movement is such that, if the fusion yields a $\Phi$, this is found later in the loop. If there are multiple possibilities for the pairing at this distance, pair with the first $\Phi$ found.

\item If there are still $\Phi$ anyons present, repeat the process for $k=k+1$.

\item When all $\Phi$ anyons have been removed, reset $k$ to $1$ and repeat for the $\Lambda$ anyons.

\end{enumerate}

Once all the anyons are removed, the total pattern of spin flips used to remove them is considered. The correction procedure is a success if this belongs to the same equivalence class as the pattern of spin flips that occurred in error. Otherwise, the correction procedure results in a logical error.

This algorithm pairs anyons that are mutual nearest neighbours. The reason why nearest neighbours are chosen is that these are more likely to have been created by the same chain of errors than more well separated anyons. However, allowing anyons to simply pair with nearest neighbours would be too greedy, since the pairing may appear likely to one of the anyons but unlikely to the other. The requirement for the pairings to be of mutual nearest neighbours then provides a barrier to such unwanted pairings.

The computational complexity of the algorithm is not deterministic. However, an upper bound for the worst case scenario can be easily derived. The maximum number of anyons present during each level of the search (i.e. each value of $k$ considered) is $O(L^2)$. For each of these, $k$ plaquettes are searched through, and so the complexity for each level is $O(k L^2)$. The maximum number of levels considered will be $O(L)$, since no anyon is more than this distance from the edge, making the total complexity never more than $O(\sum_{k=1}^{L} k L^2) = O(L^4)$. For the best case scenario, where all anyons are paired within an $O(1)$ distance, the total complexity is $O(L^2)$. Here we have neglected the $\log L$ factors required to store the necessary numbers during the process. The total complexity is clearly polynomial with system size, and a moderately low ordered polynomial also. The algorithm therefore allows for fast and efficient decoding.

\section{Results}

In order to properly benchmark the decoder, the logical error rate $P$ was determined for many spin flip error rates, $p$, and linear system sizes, $L$. In each case this was done by randomly generating an error configurations for the spins of the code according to the error rate, applying the decoder to the resulting anyons, and finally determining whether or not a logical error occurred for each sample. The number of samples, $n$, used in each case was that required in order for $10^3$ logical errors to occur. The logical error rate was then calculated as $P=10^3/n$.

\begin{figure}[H]
\begin{center}
\includegraphics[width=7.5cm]{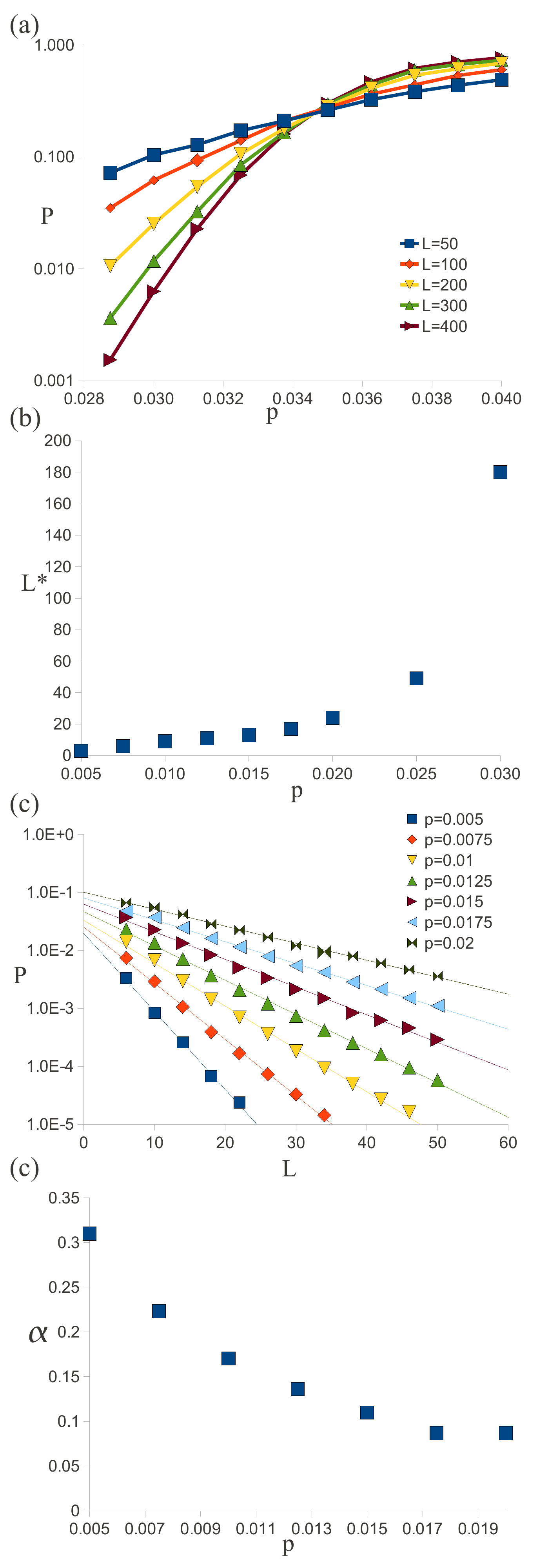}
\caption{\label{data}
(a) Graph of logical error rate, $P$, against error rates, $p$, around threshold. (b) Graph of minimum linear system size required for error correction to become evident, $L^*$, against $p$. (c) Graph of $P$ against linear system size, $L$, for various $p$ well below threshold. Fittings to an exponential of the form $P=c \, e^{-\alpha L}$ are shown. (d) A graph of $\alpha$ against $p$ calculated using this fitting. }
\end{center}
\end{figure}

The data was obtained to determine the following aspects of the algorithm's behaviour:
\begin{enumerate}[(a)]
\item The threshold error rate, $p_c$, under which error correction is possible;
\item The minimum system size required such that $P<p$, and so error correction becomes evident, for each $p$;
\item The decay of the logical error rate for spin flip error rates well below threshold, to show that effective error correction occurs;
\item The values of $\alpha(p)$ for each $p$, when the above data is fitted to a function $P = 0(e^{-\alpha(p) L})$ \cite{bravyi}.
\end{enumerate}
The results can all be found in Fig. \ref{data}. These suggest a threshold $p_c$ of around $3.5 \%$. The threshold for the total error probability on each spin will then be around $7\%$, which is very comparable to similar results from Abelian models. Indeed, the corresponding decoder and error model for the corresponding $D(Z_2)$ model also achieves a threshold of around $7 \%$ \cite{wootton13}. Note that the value of this threshold depends on our choice of decoder. The maximum possible threshold for an optimal decoder could be much greater, but will likely be of the same order of magnitude.

It is found that logical error rates of $P<p$ can be obtained using a small code of size $L<10$ for all $p \lesssim p_c/3$, and then rises sharply for higher $p$. This differs slightly to the decoding of Abelian anyons, for which such small systems sizes typically perform well up to $p_c/2$. However, this may be due to the properties of the decoder rather than the decoding problem itself. For $p \lesssim p_c/2$ we find that the logical error rate decays very quickly as $L\rightarrow \infty$.

The data shows a good fit to an exponential decay with $\beta=1$, though it is expected that a lower value of $\beta$ will become evident as $L\rightarrow \infty$. For the case of $p_{\Phi}=0$ and $p=p_{\Lambda}$ it is known that $\beta = \log_3 2$ \cite{dennis_thesis,wootton13}. We expect the same for $p_{\Phi}>0$, but this remains to be determined.

Another important benchmark of performance is the minimum number of errors required to cause a logical error. We will use $\epsilon$ to denote the value of this number realized by an exhaustive decoder, and $\epsilon'$ to denote that for a practical decoder. For this code, ignoring $O(1)$ corrections, it is clear that $\epsilon = L/2$. This is the minimum number of spin flips required to create a pair of anyons such that it takes less flips to pair them with opposite edges than with each other. This decoder, however, does not achieve this optimal behaviour. Instead, by straightforward application of the result derived for the $D(Z_2)$ planar code in \cite{wootton13}, we find $\epsilon/4 \leq \epsilon' < \epsilon/2$, up to constant $O(1)$ correction terms. This is not ideal, but the fact that it scales linearly with $L$ should allow good error suppression.

\section{Nature of syndrome extraction} \label{sec:nature}

For Abelian anyon models, the fusion product of two anyons can be predicted exactly in all cases. As such, Abelian decoding can be achieved by first measuring and then classically processing the syndrome.

The decoding process described above, however, is more complex. First the positions of anyons is measured. This gives partial syndrome information, since the fusion results of anyons remains unknown. The results are then processed to choose a pair of anyons to fuse. The result of this is measured, giving additional syndrome information. Further processing, fusion and measurement cycles are made until no more anyons are present. This decoding therefore switches constantly between partial syndrome measurement and partial processing, with the syndrome measurements being made at each step being guided by the processing of the last step. The fusion basis used to fuse all anyons and hence extract all syndrome information is therefore chosen slowly, using the results from each round of fusions to determine the nature of the next round.

It is interesting to consider whether non-Abelian decoding must necessarily take this form, or whether some means of full syndrome extraction followed by full processing is possible. In the latter case, the fusion basis used for full syndrome extraction cannot depend at all on the anyon configuration, since no processing will be performed between anyon configuration measurement and the fusion measurements in order to make this choice. The fusion basis measurement must therefore be done using a pre-determined convention, fusing the contents of the plaquettes in a certain order with no regard to what those contents are. If decoding is possible with such syndrome extractions, it should then be possible process the results of all measurements (the initial positions of the anyons as well as their fusion results) to determine which logical operation was performed by the combined error and measurement process.

To see that this is not possible, consider the creation from the vacuum of a pair of $x$ anyons for $x \in \{ 1, \Lambda, \Phi \}$. After this a pair of $\Phi$ anyons is created, with one $\Phi$ fused with each $x$. Let us consider the case that both fusions yield a $\Phi$. This always occurs for $x \in \{ 1, \Lambda \}$ and occurs with probability $1/2$ for $x=\Phi$. In all cases the result is simply a pair of $\Phi$ anyons created from vacuum, in a state that does not depend at all on the value of $x$. Indeed the fusion space of two $\Phi$'s to vacuum is one-dimensional, so there is nowhere for this information to be stored. Any $\Phi$ pair created from the vacuum is therefore equivalent to a $\Phi$ pair that fused pairwise with a randomly chosen $x$ pair created from vacuum immediately before. Equivalent statements can be made for any non-Abelian anyons of any model.

Information is stored in the code using the edge occupations. Creating a pair of anyons from vacuum and fusing one with each edge therefore performs a logical operation. Doing this randomly results in an uncorrectable logical error, since the syndrome carries no trace. By the properties of $\Phi$'s described above, fusing a $\Phi$ with each edge is equivalent to first performing an uncorrectable logical error and then fusing a $\Phi$ with each edge. As such, this process always results in an uncorrectable logical error, even when it is known that it has occurred. There is no way to reverse the fusion of a $\Phi$ with each edge to restore the original state. This property holds for any non-Abelian anyons of any model.

With this property, it is easy to see that decoding is not possible for full syndrome measurement according to predetermined fusion basis. For any such fusion basis there will be at least one spin such that, if a pair of $\Phi$'s is created by an operation on that spin, these will end up fusing with opposite edges. Only one such a pair creation is required on this single spin in order to cause an uncorrectable logical error. The probability of a logical error will then always be $\geq p_{\phi}$, and so not decay with system size at any finite error rate.

Because of this, any decoding algorithm for any non-Abelian model requires at least an initial measurement step, followed by classical processing based on the results to determine the next set of measurements to be made, followed by these measurements, followed by classical processing. Further cycles may also be required in general. This result applies for arbitrary encoding schemes, not just those using the edge.

\section{Continuous error correction}

The error model considered in this work is one for the one-time case, where the system is first prepared, then errors occur, and then (perfect) measurement and processing is performed to remove the effects of the errors at time of readout. The relevant timescale between preparation and readout is then $O(1)$ (unless a self-correcting Hamiltonian such as \cite{wootton_rev,fabio} is applied). Not considered is the case for which errors occur continuously. In this case the aim of error correction is to allow the effects all errors to be removed over a typical timescale of $O(\exp[O(L^\beta)])$.

For the decoding of Abelian anyons using perfect syndrome measurements, the existence of a decoding algorithm for the one-time case that achieves exponential suppression of the logical error rate directly implies the existence of a decoding algorithm for the continuous case that achieves an exponential lifetime. This is because the code can simply have a full syndrome measurement performed periodically (with $O(1)$ time between each) until the time of readout. Processing can be done as the measurements are made, or can be deferred until the end. For each time slice, the syndrome is taken to be the difference between the present measurement result of the previous one, giving only the effects of the errors that occurred between the two. The decoding can then be performed independently for each time slice. The cumulative results of these allow the total logical operation performed on the stored information to be determined and its effects negated. As long as the error rate between time slices remains below the threshold value for the one-time decoder, the probability that the decoding fails at each time slice decays exponentially with $L^\beta$. The typical time before such a logical error occurs, the lifetime of the quantum memory stored in the code, is then $O(\exp[O(L^\beta)])$.

The above does not apply to non-Abelian decoding due to the behaviour described in the previous section. One difference in the the non-Abelian case is that processing cannot be deferred until the end. As soon as the probability of an error that creates a $\Phi$ pair (or any particle-antiparticle pair of non-Abelian anyons in any non-Abelian model) on each spin is above the bond percolation threshold of the lattice, which will occur in $O(1)$ time, the effect will be the same as a pair of $\Phi$'s for which each is fused with a different edge. Even though the periodic measurements may allow the decoder to know such a process has occurred, it will nevertheless cause an uncorrectable logical error. Constant operations are therefore required to keep the number of anyons sufficiently low.

Another difference is that full decoding cannot be performed based on the results of a single round of measurements.  The decoder must instead make measurements, process the results, and then make more measurements based on these. However, if further errors occur during the time taken by this processing, the exact nature of the measurements to be made becomes ambiguous. The processing could have called for two anyons to be fused and the result measured. However, if they move during the time taken to determine this, it is no longer clear where those anyons are. They may also have fused with other anyons. The decoder could therefore do more processing to determine how to change its plans accordingly. However, in this time more errors and so more changes will have occurred.

Another possibility is to attempt the operations specified by the processing without regard to any changes. This would be possible in the Abelian case, where the operation required to fuse any two anyons is a well defined product of single spin flips. This can therefore be performed even if the errors have moved the anyons away. The effect will be the same as if the anyons were fused first and the errors followed after. However, non-Abelian anyons can only be moved using controlled operations. Clearly, if the anyon is no longer present in its expected position, the fusion required by the processing cannot be performed.

Due to the above complications, the decoding algorithm used for the above numerics only applies in the case of one-time error correction. The lifetime of the quantum memory in this case would therefore need to be extended using additional techniques, such as energetic suppression of errors or a self-correcting Hamiltonian \cite{wootton_rev,fabio}. The problem of how decoding can be achieved for continuous error correction is still open.

Note that the above considerations also apply to the case of imperfect syndrome measurements, where many syndrome measurements must be made in order to accurately determine the true syndrome. Because of this, even one-time readout will need some aspects of continuous decoding.

\section{Conclusions}

We have shown that good decoding, which achieves logical error rates suppressed exponentially with $L$ below a finite threshold noise rate, is indeed possible for the $\Phi-\Lambda$ non-Abelian anyon model for a simple case of noise. This result can be expected to hold for many other, if not all, non-Abelian decoding problems based on similar noise models.

We also showed that, unlike the Abelian case, the proof of principle for this simple one-time noise model cannot be straightforwardly extended to the more general case of continuous error correction. No proof of principle for this case has yet been made, especially with the more realistic consideration of imperfect syndrome measurements.

Future work is to progress towards the decoding of the model for the full fault-tolerant case, consisting of continuous error correction and imperfect syndrome measurements. This should determine whether or not there exists a decoding algorithm with computationally efficient processing that is able to provide such error correction. If so, it would provide a very good justification that such decoding algorithms also exist for Fibonacci anyons and other universal models.

Note that, while our work was in preparation, the problem of decoding the Ising model was discussed in \cite{talk}.

\section{Acknowledgements}

The authors would like to thank Ville Lahtinen for discussions and the Swiss NF, NCCR Nano and NCCR QSIT for support.

\appendix

\section{Fusion and braiding in the $\Phi-\Lambda$ model} \label{app:phi-lambda}

To demonstrate the equivalence of the classical simulation and the $\Phi-\Lambda$ model, we will explicitly consider the case of four $\Phi$ pairs created from the vacuum. This case contains all forms of behaviour present in more general anyon configurations. The equivalence of the classical simulation and the $\Phi-\Lambda$ model in this case is therefore strong evidence for the equivalence in general (A full proof can be found in Appendix \ref{app:DS3}). The relationship between braiding and permutation, the way this is represented as an action on the fusion space is also discussed.

\subsection{Fusion and braiding of four $\Phi$ anyons}

We consider in detail the behaviour of the $\Phi-\Lambda$ model for the simplest non-trivial case, that of four $\Phi$'s created from the vacuum. There are only two possible fusion bases in this case, as depicted in Fig. \ref{F&R} (b). Assuming that only topological operations (splitting, braiding and fusion) are allowed, the statistics of the model are fully described by the statistics of fusion outcomes.

First, we consider the case that no braiding occurs between creation and fusion. If the anyons are fused in the same basis as which they were created, the intermediate particle found in the fusion measurement will be exactly the same as that appearing in the creation. When fusion is done in a different basis to creation, there will be some randomness to the results. We use $p(x' | x)$ to denote the probability that, when the state is created according to the left basis (in Fig. \ref{F&R} (b)) with $x$ used in the intermediate step, fusion according to the right basis will yield $x'$ in the intermediate step. Using the $F$ matrix of Eq. (\ref{ChargeF}), we find that these probabilities are,
\be \label{Probs}
p(x'|x) = \frac{1}{4}, \,\,\, p(x'|\Phi)= p(\Phi | x) = \frac{1}{2}, \,\,\, p(\Phi | \Phi) = 0
\ee
where $x$ and $x'$ in the above are restricted to the set $\{1, \Lambda \}$. Note that, since the $F$ matrix is Hermitian, these probabilities also apply when states prepared using the left basis are fused using the right.

We now consider the case for which braiding can occur. Without loss of generality we can assume all braiding occurs after the creation of the all the particles and before any fusion. If any two anyons with a definite fusion outcome are exchanged, the effect is trivial. We therefore ignore these. Since braiding of the $\Phi$ anyons is a representation of the permutation group, it is clear that any two successive non-trivial exchanges will return the anyons to their initial configuration, or will map between the fusion bases. In either case, the state prior to fusion is simply a fusion basis state and so is described by the probabilities above. It is therefore sufficient to consider only the effects of single exchanges.

For a state prepared in the left basis, the only non-trivial exchange is that of the two middle anyons. For a state of the right basis there are two possible non-trivial exchanges that are equivalent in effect. These are of the left pair of anyons or of the right pair. In all cases, the resulting state yields the same probabilities as Eq. (\ref{Probs}) for fusion in both bases (unlike before, neither basis now has a deterministic result).

Given that braiding only represents permutation, and has no additional effect, these probabilities can be expressed more simply as follows. For either a left or right basis state with intermediate using $x$, we can think of the four anyons as being two pairs, each split from an anyon of type $x$. The probabilities for outcomes when fusing anyons from the same pair, whether braiding has been used to move these towards each other or not, is $p(x'|x) = \delta_{x,x'}$. The probabilities when fusing from different pairs (whether braiding is used or not) are those of Eq. (\ref{Probs}).

These can now be compared to the probabilities for fusion outcomes in the classical simulation. From Eq. (\ref{create_probs}) it is clear that the state of the two pairs for $x \in \{1, \Lambda \}$ is
\be
\frac{1}{16} \sum_{j,k \in \phi} \ket{ \beta_j , \beta_{X-j}, \beta_k , \beta_{X-k} } \bra{ \beta_j , \beta_{X-k}, \beta_k , \beta_{X-j} }.
\ee
Here $X=0$ for $x=1$ and $X=3$ for $x=\Lambda$. Clearly the internal states for two $\Phi$ anyons from different pairs are uncorrelated. Their reduced density matrix will simply be $\rho_{\Phi}^{\otimes 2}$, where
\be
\rho_{\Phi} = \sum_{j \in \phi} \ket{ \beta_j} \bra {\beta_j}.
\ee
Since fusion to vacuum corresponds to four possibilities ($k=-j$ for $j \in \phi$) as does fusion to a $\Lambda$ ($k=3-j$ for $j \in \phi$), whereas $\Phi$ corresponds to eight possibilities ($k=j$ and $k=j+3$ for $j \in \phi$), the probabilities for the fusion results will be as in Eq. (\ref{Probs}) for the $x \in \{1, \Lambda \}$ case.

For the case of $x = \Phi$ the state of the two pairs will be
\bq \nonumber
\frac{1}{16} \sum_{i \in \{1,2\}, \atop j,k,l \in \{0,3\}} &&  \ket{ \beta_{i+j} , \beta_{i+k}, \beta_{-i+l}, \beta_{-i+j+k+l} } \\
&& \bra{  \beta_{i+j} , \beta_{i+k}, \beta_{-i+l}, \beta_{-i+j+k+l} }
\eq
The reduced density matrix of two anyons from different pairs is then
\be
\frac{1}{8}\sum_{i \in \{1,2\}, \atop k,l \in \{0,3\}} \ket{\beta_{i+k}, \beta_{-i+l}} \bra {\beta_{i+k}, \beta_{-i+l}}.
\ee
The fusion outcome will be $\beta_{k+l}$ for $k,l \in \{0,3\}$. The only two possible outcomes are then vacuum and $\Lambda$, which each occur with probability $1/2$. This reproduces the probabilities of Eq. (\ref{Probs}) for the $x = \Phi$ case. The classical simulation is therefore statistically equivalent to the case of four $\Phi$ anyons created from vacuum in all cases

\subsection{Braiding and permutation}

The braiding of this model only represents the permutation of the anyons, and has no additional effect on the fusion space. However, it is important to note that this does not mean the effect on the fusion space is trivial. The abstract theory of anyons treats anyons of the same type as indistinguishable. Different states of these anyons are therefore only distinguished by their fusion basis states, which describe the outcomes of fusing anyons in neighbouring positions. Permutation of anyons changes which anyons are in which positions. A state for which two neighbouring positions are occupied by anyons with a definite fusion outcome might then be changed to one for which the anyons in these positions have a random outcome. This would clearly change the state of the system, and therefore must act non-trivially on the fusion space.

It is also important to note that the braiding is a non-Abelian representation of the permutation group. As an example of this, consider the case of four $\Phi$ anyons as above. Specifically, consider the left basis state with $x=1$. Let us label the positions of the anyons $A$ to $D$ from left to right, and also label the anyons themselves $\Phi_a$ to $\Phi_d$ from left to right. The initial state is then one for which each $\Phi_j$ is in position $J$, and $\Phi_a \times \Phi_b=\Phi_c \times \Phi_d=1$.

Now let us consider two exchange operations: (i) the exchange of the anyons at positions $A$ and $B$ and (ii) that for the anyons at $B$ and $C$. After both are complete, the anyons in positions $B$ and $C$ will be fused. If exchange (i) is applied before (ii), the anyons fused at the end will be $\Phi_a$ and $\Phi_c$. These are from different pairs, and so the result will be random. If the exchange (ii) is applied before (i) the anyons fused are $\Phi_a$ and $\Phi_b$. Since the braiding has no effect beyond permutation, the $\Phi_a \times \Phi_b=1$ behaviour from the initial state is retained. The outcome of the fusion is then always vacuum. Since the probability distribution for the fusion outcome depends on the order of exchanges, it is clear that the braiding is non-Abelian.

\section{Classical simulation of $D(S_3)$ lattice model subject to noise} \label{app:DS3}

The noise experienced by topological codes based on non-Abelian anyons will depend on various factors, such as the details of the physical system used and the nature of the coupling between system and environment. Nevertheless, the noise will always be interpreted in terms of the creation, transport and fusion of the anyons. Since our aim in this study is to consider the decoding problem from a general anyonic perspective, we therefore chose an error model that is straightforward in terms of these simple anyonic processes. To do this for the $\Phi-\Lambda$ model, let us first look at the physical system on which it may be realized.

The $D(S_3)$ lattice model is the standard lattice realization of the $\Phi$ and $\Lambda$ anyons. This can be found using the construction of \cite{double} with the group $S_3$. This model was considered in greater detail in \cite{brennen}. The group $S_3$ is the permutation group of three objects, which has two generators $t$ and $c$. Using the rules $t^2=c^3=e$ and $tc=c^2t$ it is clear that the group has six elements that can be denoted $e$, $c$, $c^2$, $t$, $tc$ and $tc^2$.

Like the lattice model used in Section \ref{sec:classical}, the model is defined on a square lattice with a six level spin on each edge. However in this case the basis states are labelled by the elements of $S_3$. Both vertices and plaquettes can hold quasiparticles, and so corresponding projectors are defined on each. Charge anyons reside on plaquettes, and so-called flux anyons reside on vertices \cite{note}. Since we are interested only in the charges, we consider the case of vacuum on each vertex. The simplest state to satisfy this is that for which all spins are in state $\ket{e}$. We will therefore use this as our starting point when defining charge states.

We define operators on the spins according to the group multiplication. We define both right and left multiplication operators $R^h \ket{g} = \ket{gh}$ and $L^h \ket{g} = \ket{hg}$, respectively, for $g,h \in S_3$. With these we define the gauge transformations,
\be
T_g = \prod_{j \in p} R^g_j, \,\,\,\,\,\, T_g = \prod_{j \in p} L^{g^{-1}}_j,
\ee
on white and grey plaquettes, respectively. These can then be used to construct the following projectors for each quasiparticle type,
\bq \label{eq:proj}
P_1 &=& \frac{1}{6}\left( T_e + T_c + T_{c^2} + T_{t} + T_{tc} + T_{tc^2} \right), \\
P_{\Lambda} &=& \frac{1}{6}\left( T_e + T_c + T_{c^2} - T_{t} - T_{tc} - T_{tc^2} \right), \\
P_{\phi} &=& \frac{1}{3} \left( T_e + \omega T_c + \omega^2 T_{c^2} \right), \\
P_{\bar\phi} &=& \frac{1}{3} \left( T_e + \omega^2 T_c + \omega T_{c^2} \right).
\eq
Here $\omega = e^{i 2 \pi/3}$. These projectors are mutually orthogonal, and commute with each other and the vertex projectors. The projector $P_1$ corresponds to states for which the plaquette holds vacuum charge, and $P_{\Lambda}$ corresponds to the $\Lambda$ anyon. The projectors $P_{\phi}$ and $P_{\bar\phi}$ correspond to different internal states of the $\Phi$ anyon. A single $\Phi$ projector can be obtained from their sum,
\be
P_{\Phi} = \frac{1}{3} \left( 2T_e - T_c - T_{c^2}  \right).
\ee
However, it will be advantageous in the following to consider the internal states separately. We will refer to these as `quasiparticles' rather than `anyons'.

The $\phi$ and $\bar\phi$ quasiparticles can be considered to be antiparticles of each other. To create a $\phi$ on a white plaquette and a $\bar\phi$ on a neighboring grey plaquette, the operator
\be
W_{\phi} = \ket{e}\bra{e} + \omega \ket{c}\bra{c} + \omega^2 \ket{c^2}\bra{c^2}
\ee
should be applied to the shared spin. To create the $\phi$ on the grey plaquette and $\bar\phi$ on the white, the operator $W_{\bar\phi}  = W_{\phi}^\dagger$ should be applied. A $\Phi$ pair corresponds to a superposition of these two possibilities, and is created with $W_{\Phi} = W_{\phi}-W_{\bar\phi}$ \cite{wootton_ds3,brennen}.

Note that these operators are not unitary, and so the state after their application will need to be renormalized. However, they can be applied deterministically using adaptive operations \cite{brennen}.

A pair of $\Lambda$ anyons can similarly be created by the unitary operator 
\bq \nonumber
W_{\Lambda} &=& \ket{e}\bra{e} + \ket{c}\bra{c} +  \ket{c^2}\bra{c^2} \\
&-&  \ket{t}\bra{t} - \ket{tc}\bra{tc} -  \ket{tc^2}\bra{tc^2}
\eq
Note that all these creation operators are diagonal in the basis labelled by group elements. This ensures that they do not create, annihilate or transport any quasiparticles present on the vertices \cite{brennen}.

These operators not only create anyons on initially empty plaquettes, but will also fuse the anyons with whatever is initially present. They therefore implement all the basic anyonic operations of creation, transport and fusion. The simplest error model will then be one for which $W_{\Phi}$ or $W_{\Lambda}$ is applied to each spin with respective probabilities $p_{\Phi}$ and $p_{\Lambda}$. We will now show that this error model can be classically simulated as described in Sections \ref{sec:classical} and \ref{sec:error}.

To do this we will consider the creation of quasiparticle pairs such that they overlap on a plaquette $p$. By doing so, we will be able to determine their fusion behaviour by applying the projectors of Eq. (\ref{eq:proj}) to the resulting state on $p$. In order to ensure that the results are due only to the quasiparticles created, and not to any pre-existing ones, the initial state we will use will be such that the plaquette $p$ and all four surrounding plaquettes hold only vacuum, as do all vertices. As noted above, all vertices hold vacuum when all spins are in state $\ket{e}$. We may then apply the $P_1$ projectors for $p$ and the surrounding plaquettes to obtain a state for which these also hold only vacuum. Since we are only interested in the outcome of the fusion on $p$, we then take the reduced density matrix for the spins of $p$. This yields an equally weighted mixture of all states of the form
\be
\frac{1}{\sqrt{6}} \sum_{g \in S_3} \ket{g,g h_2, g h_3, g h_4}.
\ee
for $h_j \in S_3$. These states apply to both white and grey $p$. Note that the summation variable $g$ is used for the state of the first spin, and the state of the $j$th spin differs by a relative factor $h_j$. However, this is a notational convenience and does not confer any special status on the first spin.

Since these states correspond to vacuum on plaquette $p$, we will denote them $\ket{ 1_{h_2,h_3,h_4} }$. We will also use the notation $\ket{G_{h_2,h_3,h_4}} = \ket{g,g h_2, g h_3, g h_4}$ These states may then be expressed
\be
\ket{1_{h_2,h_3,h_4}} = \frac{1}{\sqrt{6}} \sum_{g \in S_3} \ket{G_{h_2,h_3,h_4}}.
\ee
It can be easily verified that $P_1 \ket{ 1_{h_2,h_3,h_4} } = \ket{ 1_{h_2,h_3,h_4} }$ and $P_{\Lambda} \ket{ 1_{h_2,h_3,h_4} } = P_{\phi} \ket{ 1_{h_2,h_3,h_4} } = P_{\bar\phi} \ket{ 1_{h_2,h_3,h_4} } = 0$ for all such states.

Since we have a mixture of states, which prohibits interference effects between them, we can consider each separately. We will first apply the operator $W_{\Lambda}$ to the first spin (though the effect on any other would be equivalent). This yields
\be \nonumber
\ket{\Lambda_{h_2,h_3,h_4}} = \frac{1}{\sqrt{3}} \left( \sum_{g \notin [t]} \ket{G_{h_2,h_3,h_4}}
- \sum_{g \in [t]} \ket{G_{h_2,h_3,h_4}} \right),
\ee
where $[t] = \{t, tc, tc^2\}$. It can be easily verified that $P_1 \ket{\Lambda} = P_{\phi} \ket{\Lambda} = P_{\bar\phi} \ket{\Lambda} = 0$ and $P_{\Lambda} \ket{\Lambda} = \ket{\Lambda}$ for both grey and white plaquettes. So the application of $W_{\Lambda}$ to a spin creates a $\Lambda$ on both adjacent plaquettes. Applying  $W_{\Lambda}$ twice on the same plaquette, either on the same spin or on different spins, $i$ and $j$, clearly gives $W_{\Lambda}^i W_{\Lambda}^j \ket{1} = \pm \ket{1}$. The global phase of $-1$ occurs if $h_i$ belongs to $[t]$ and $h_j$ does not, or vice-versa. However, as a global phase to a state within a mixture, it has no physical effects. Fusing two $\Lambda$ anyons by placing them on the same plaquette therefore always leads to their annihilation, and thus realizes the $\Lambda \times \Lambda = 1$ fusion rule.

Now we consider the $\phi$ and $\bar\phi$ quasiparticles. Applying the operator $W_{\phi}$ to the first spin (though the effect on any other would be equivalent) for the vacuum state $\ket{1_{h_2,h_3,h_4}}$ yields
\bq \nonumber
\ket{\phi_{h_2,h_3,h_4}} &=& \frac{1}{\sqrt{3}} ( \ket{E_{h_2,h_3,h_4}} + \omega \ket{C_{h_2,h_3,h_4}} \\ \nonumber
&+&  \omega^2 \ket{C^2_{h_2,h_3,h_4}} )
\eq
It is easy to see that $P_1 \ket{\phi_{h_2,h_3,h_4}} = P_{\Lambda} \ket{\phi_{h_2,h_3,h_4}} = 0$ for both the case that the plaquette $p$ is a grey plaquette and the case that it is a white plaquette. In the former case, the $\Phi$ projectors act as $P_{\phi } \ket{\phi_{h_2,h_3,h_4}} = \ket{\phi_{h_2,h_3,h_4}}$ and $P_{\bar\phi } \ket{\phi_{h_2,h_3,h_4}} = 0$. In the latter case the roles of $\phi$ and $\bar\phi$ are reversed. This shows that the effect of the $W_{\phi}$ operator on a spin is to create a $\phi$ on the white plaquette adjacent to the spin and a $\bar\phi$ on the grey plaquette.

The operator $W_{\bar\phi}$ applied to the vacuum state yields,
\bq \nonumber
\ket{\bar\phi_{h_2,h_3,h_4}} &=& \frac{1}{\sqrt{3}} ( \ket{E_{h_2,h_3,h_4}} + \omega^2 \ket{C_{h_2,h_3,h_4}} \\ \nonumber
&+& \omega \ket{C^2_{h_2,h_3,h_4}} )
\eq
This behaves similarly to the state $\ket{\phi_{h_2,h_3,h_4}}$. The only difference is that here the $\bar\phi$ is placed on the white plaquette and the $\phi$ on the grey.

Let us now consider the effect of applying the operator $W_{\phi}$ (or equivalently $W_{\bar\phi}$) twice to the same plaquette. This corresponds either moving two $\phi$ quasiparticles into the same white plaquette or two $\bar\phi$ quasiparticles into the same grey. In both cases, this results in the fusion of the quasiparticles. If both operators are applied to the same spin we may use the fact that $(W_{\phi}^i)^2 = W_{\bar\phi}$. The fusion of two $\phi$ quasiparticles on the same plaquette will then deterministically yield a $\bar\phi$, and vice-versa, in this case.

If, however, one $W_{\phi}$ is applied to the first spin and the other to the $j$th the effect depends on the relative factor $h_j$ according to
\bq \nonumber
W_{\phi}^1 W_{\phi}^j \ket{1_{h_2,h_3,h_4}} &=& \Omega(h_j) \ket{\bar\phi_{h_2,h_3,h_4}}, \\ \nonumber
\Omega(c^n) = \omega^n, \,&&\, \Omega(g \in [t]) = 0.
\eq
The state is then either projected out, the effects of which will be dealt with when the mixed state as a whole is renormalized, or yields $\ket{\bar\phi_{h_2,h_3,h_4}}$ up to an irrelevant global phase. The final mixed state will then be composed only of states for which a $\bar\phi$ resides on $p$. This case therefore also results in two $\phi$ quasiparticles fusing deterministically to a $\bar\phi$, and vice-versa. This fusion behaviour can therefore be represented with the fusion rules $\phi \times \phi = \bar \phi$ and $\bar \phi \times \bar \phi = \phi$.

The fusion behaviour of a $\phi$ with a $\bar\phi$ can be determined by applying both $W_{\phi}$ and $W_{\bar\phi}$ to the vacuum state. This yields
\be
W_{\bar\phi}^1 W_{\phi}^j \ket{1_{h_2,h_3,h_4}} = \Omega(h_j) (\openone + W_{\Lambda}^1)\ket{1_{h_2,h_3,h_4}}.
\ee
The result of the fusion is then randomly either vacuum or $\Lambda$, with an irrelevant global phase. This fusion behaviour can be represented with the fusion rule $\phi \times \bar\phi = 1 + \Lambda$.

Finally, let us consider creating a $\Phi$ (either in the form of a $\phi$ or $\bar\phi$) on the same plaquette as a $\Lambda$. It can easily be seen that
\bq\nonumber
W_{\phi}^i W_{\Lambda}^j \ket{1} = \pm W_{\phi}^i \ket{1_{h_2,h_3,h_4}} = \pm\ket{\phi_{h_2,h_3,h_4}}, \\ \nonumber
W_{\bar\phi}^i W_{\Lambda}^j \ket{1} = \pm W_{\phi}^i \ket{1_{h_2,h_3,h_4}} = \pm \ket{\bar\phi_{h_2,h_3,h_4}}.
\eq
The $W_{\Lambda}$ therefore has no effect (up to an irrelevant global phase) when applied in conjunction with the $W_{\phi}$ or $W_{\bar\phi}$, and so the $\Lambda$ is hidden. This can be represented by the fusion rules $\phi \times \Lambda = \phi$ and $\bar\phi \times \Lambda = \bar\phi$.

The process by which quasiparticles may be moved from one plaquette to a neighbouring plaquette is as follows. Firstly, an operator is applied to the spin shared by both the initial and final plaquettes. This should be the operator that creates the antiparticle of the quasiparticle to be moved on the initial plaquette, and the quasiparticle on the final plaquette. For the case of $\Lambda$ anyons the operation is then complete, since the resulting occupation of the initial plaquette is vacuum. When moving a $\phi$ or $\bar\phi$, however, the occupation of the initial plaquette will be either vacuum or $\Lambda$. A measurement should then be made to determine which is the case. If a $\Lambda$ is present, $W_{\Lambda}$ should be applied to move this onto the final plaquette into the $\phi$ or $\bar\phi$ that was moved. This then completes the transport of the quasiparticle. As noted in \cite{brennen}, the operators that create and move the quasiparticles are diagonal in the basis labelled by group elements. As such they trivially commute, and so the braiding of the  quasiparticles has no effect on the fusion beyond permutation.

Now we have determined the fusion and braiding behaviour of the $\phi$, $\bar\phi$ and $\Lambda$ quasiparticles, we can determine the statistics of measurement outcomes when a given pattern of the $W_{\Lambda}$, $W_{\phi}$ and $W_{\bar\phi}$ operations are applied to the spins of the model. Given the fusion behaviour it is clear that a definite pattern of the operators will lead to a definite pattern of $\phi$ and $\bar\phi$ quasiparticles. However, any plaquette on which a $\phi$ and $\bar\phi$ fused will lead to randomness in the result, with both vacuum and $\Lambda$ being possible. For a given such measurement, both outcomes will occur with equal probability. However, we can also consider what correlations may be present between the outcomes. The conservation law for the anyons means that, once all $\phi$ and $\bar\phi$ quasiparticles are fused the end result must be vacuum. There is therefore an overall parity constraint on the outcomes of these fusions. This is in fact the only correlation present, with the outcomes being otherwise completely random. This can be seen from the fact that the quasiparticle creation operators satisfy $ W_{\Lambda} W_{\phi} = W_{\phi}$ and $W_{\Lambda}  W_{\bar\phi} = W_{\bar\phi}$. Since the application of $W_{\Lambda}$ to any spin on which  $W_{\phi}$ or $W_{\bar\phi}$ have already been applied yields no effect, one can randomly apply $W_{\Lambda}$ with probability $1/2$ to all spins on which these operations have been applied. Such a process would clearly randomize the outcomes of any fusions (while respecting the parity constraint) and so would wash out any additional correlations present. Since the process has no effect, it follows that no such correlations are present.

In the above we consider the $W_{\phi}$ and $W_{\bar\phi}$ operators and the $\phi$ and $\bar\phi$ quasiparticles that they create. However, our real interest is in $\Phi$ anyons created by $W_{\Phi} = W_{\phi}-W_{\bar\phi}$. Note that any application of $W_{\Phi} = W_{\phi}-W_{\bar\phi}$ on a spin creates an equally weighted superposition of the states that would result from applications of $W_{\phi}$ and $W_{\bar\phi}$. The probability distribution for measurement outcomes after a pattern of $W_{\Phi}$ operations are applied is then similar to that if each $W_{\Phi}$ was replaced by either $W_{\phi}$ or $W_{\bar\phi}$ with equal probability. The difference is that the probabilities for the former superposition state will be subject to interference effects whereas those for the latter mixed state will not. However, a regime can be easily defined for which there are no interference effects, and so probability distributions for measurement outcomes in the two cases are equivalent.

This regime is that for which the superpositions caused by the fusion of two $\Phi$ anyons are immediately decohered. This could be due to a measurement of the corresponding plaquette occupancy immediately after the fusion. Note that such a measurement will be performed anyway by the syndrome measurement. The requirement for decoherence is then only relevant in the case for which two $\Phi$ anyons fuse on a plaquette, and then at least one more $\Phi$ fuses with the result. With the error model considered in this work, where the probability of a $\Phi$ creation operation applied to any spin is $\sim 1 \%$, the probability of more than two such errors on any plaquette is $\sim 10^{-4} \%$. Since the need to perform the decoherence is so unlikely, it seems reasonable that it will not have a significant effect on the nature our results. Furthermore, known means by which the anyons can be created and moved uses an ancilla assisted adaptive procedure. This naturally causes decoherence of the fusion result \cite{brennen}. The means by which the such operations are applied by the environment can then also be expected to cause such decoherence. We therefore feel that the study of this regime is well justified.

To see that this regime does not allow interference effects to change the probabilities when a given pattern of $W_{\Lambda}$ and $W_{\Phi}$ operations are applied to the spins, consider the application of the operations one-by-one according to some order. Physically, this will be the temporal order in which the operations were applied first by the environment and then by the error correction process. To discuss intermediate states, let us use $n$ to be the number of operations applied so far, with any decoherence applied as required during the process. If the $n+1$th operation to be applied is $W_{\Phi}$ on the spin shared by two plaquettes $p$ and $p'$, there are four distinct cases to consider. 

The first is that for which the state has no $\Phi$ on either $p$ nor $p'$. The $W_{\Phi}$ will then simply create a $\Phi$ on both. These will absorb any $\Lambda$ anyons initially present on the plaquettes. Since the decoherence ensures that any randomness in the $\Lambda$ occupation is not due to a coherent superposition, such fusion does not lead to any unwanted coherence in the fusion space.

The second case is that for which the state after the first $n$ operations is such that $p$ holds a $\Phi$,but $p'$ does not (or vice-versa). The result on $p$ will depend on the internal state of the two particles fused. A $\Phi$ will result, for example, if both have the same internal state $\phi$ or $\bar\phi$. Since these two internal states yield the same result, one might expect that the probability for that result could depend on interference effects between them. To show that this is not the case, the state of the system can be expressed in the basis of anyon occupancies as
\be
\frac{1}{\sqrt{2}} \left( \ket{\phi}_p \otimes \ket{1}_{p'} \otimes \ket{\bar\phi}_{rest} - \ket{\bar\phi}_p \otimes \ket{1}_{p'} \otimes \ket{\phi}_{rest} \right)
\ee
Note that $\ket{\phi}_p$ and $\ket{\phi}_p$ here represent the state of the plaquettes $p$ and $p'$ holding quasiparticle $\phi$ and vacuum, respectively. They do not represent the state of the four spins around $p$, as was used earlier. Also $\ket{\bar\phi}_{rest}$ denotes the state of all other plaquettes and the edges. The exact details of this need not be known. The only important point is that the fusion of all such quasiparticles will yield a $\Phi$ with opposite internal state to that on $p$, such that the two fuse to vacuum. Due to this entanglement, the state of the occupancy of $p$ alone can be expressed (in the basis of anyon occupancies) as
\be \nonumber
\rho_p = \frac{1}{2} \left( \ket{\phi}_p\bra{\phi} + \ket{\bar\phi}_p\bra{\bar\phi} \right)
\ee
Since this is a mixture of the two internal states and has no coherence, it is clear that the probabilities for outcomes when applying $W_{\Phi}$ will not be affected by interference effects.

The third case is that both $p$ and $p'$ hold a $\Phi$, created by $W_{\Phi}$ operations on spins other than that shared by the two. The $\Phi$'s with which these were created in pairs reside in the rest of the lattice, and so ensure that the rest of the lattice knows the internal state of both. This entanglement again ensures that the mixed state of the plaquettes $p$ and $p'$ does not have sufficient coherence to allow interference effects when the $W_{\Phi}$ is applied.

Because of the above behaviour, an error model that randomly creates $\Lambda$ and $\Phi$ anyons by applying $\openone$,  $W_{\Lambda}$ or $W_{\Phi}$ independently to each spin with respective probabilities $1-p_{\Phi}-p_{\Lambda}$, $p_{\Lambda}$ and $p_{\Phi}$ gives the same probability distribution for measurement outcomes as one which applies $\openone$,  $W_{\Lambda}$, $W_{\phi}$ or $W_{\bar\phi}$ with respective probabilities $1-p_{\Phi}-p_{\Lambda}$, $p_{\Lambda}$, $p_{\Phi}/2$ and $p_{\Phi}/2$. It now remains to show that this can be realized by the classical model used in our study.

In the classical model of Section \ref{sec:classical} we associate $\Lambda$ anyons with $\beta = 3$ and $\Phi$ anyons with $\beta \in \{1,2,4,5\}$. For the quasiparticles discussed here we split the latter into $\phi$ quasiparticles for $\beta \in \{1,4\}$ and $\bar\phi$ quasiparticles for  $\beta \in \{2,5\}$. The result of fusing two quasiparticles with corresponding internal states $\beta_i$ and $\beta_j$ is $\beta_k = \beta_i + \beta_j \mod 6$. From this it is evident that the classical model reproduces the fusion rules found above for $\phi$ and $\bar\phi$.

Pairs of $\Lambda$ anyons are created in the classical model using the operator $R^3$. This therefore plays the same role as $W_{\Lambda}$. The $W_{\phi}$ operator in the $D(S_3)$ lattice model creates a $\phi$ on a white plaquette and a $\bar\phi$ on the neighbouring grey plaquette. For the classical model the same can be achieved by applying the $R^1$ or $R4$ operators. However, note that though $W_{\Lambda} W_{\phi} = W_{\phi}$, and so the creation of a $\Lambda$ pair on top of a $\phi / \bar\phi$ pair has no effect, the same is not true for the classical operators. Instead $R^3 R^1 = R^4$ and $R^3 R^4 = R^1$. This can be rectified by not creating $\phi / \bar\phi$ pair using a definite choice of either $R^1$ or $R^4$, but instead using a random choice. If the probability of choosing each is equal, the permutation of the two caused by the $R^3$ has no effect. The $W_{\bar\phi}$ operation similarly corresponds to the random application of $R^2$ or $R^5$. The operator $W_{\Phi}$ then corresponds to the random application of $R^1$, $R^2$, $R^4$ or $R^5$, all with equal probability. The error model considered in Section \ref{sec:error} for the classical model is therefore equivalent to that on the $D(S_3)$ lattice model where $\openone$,  $W_{\Lambda}$ or $W_{\Phi}$ are applied independently to each spin with respective probabilities $1-p_{\Phi}-p_{\Lambda}$, $p_{\Lambda}$ and $p_{\Phi}$.


Also note that the decoding problem of the main text assumes that the syndrome measurements correspond to the projectors $P_1$, $P_\Lambda$ and $P_{\Phi}$. However, one could instead replace the latter with $P_{\phi}$ and $P_{\bar\phi}$, giving more detailed syndrome information. This case is not considered due to the fact that this trick is specific to this model, and the aim of this study is to consider general behaviour as much as possible. However, it is interesting to ask whether measurements which allow such greater detail to be extracted may be present for other non-Abelian models.

\section{Applicability of the decoder to the Fibonacci anyon model} \label{app:fibonacci}

The Fibonacci model consists of a single (non-trivial) anyon type, $\tau$, with fusion rule
$\tau \times \tau = 1 + \tau$.
For a decoding algorithm to be applicable to a syndrome that consists of such anyons it therefore:
\begin{enumerate}
\item must be able to deal with the possibility of a pair of non-Abelian anyons fusing to a non-Abelian anyon (since this can happen in the Fibonacci model);
\item does not need to distinguish between different types of non-Abelian anyon in order to gain enough information to decode (since the model has only one type);
\item does not need to use information regarding any Abelian anyons to decode (since the model has none).
\end{enumerate}
The $\Phi-\Lambda$ model has two non-trivial anyon types: a non-Abelian anyon $\Phi$ and an abelian anyon $\Lambda$. The most important fusion rule is $\Phi \times \Phi = 1 + \Lambda + \Phi$. From this we can see that a decoding algorithm applicable to these anyons must also fulfil the first two requirements above, since two $\Phi$ anyons can fuse to a $\Phi$ and since this is the only type of non-Abelian anyon. However, such a decoder need not fulfil the third. Instead, it could use information concerning the presence of $\Lambda$ anyons to better decode the $\Phi$'s. This possibility is due to the fact that the creation of two $\Phi$ anyons on the same plaquette will result in a $\Phi \times \Phi$ fusion, which can yield a $\Lambda$. Since $\Lambda$'s can be created from the same errors that create $\Phi$'s by this secondary process, the $\Lambda$ syndrome will be helpful in correcting these errors.

Despite this possible advantage, the decoder we consider does not use the $\Lambda$ syndrome to help decode the $\Phi$ anyons. Instead it first runs a process that considers the $\Phi$ anyons alone, and then considers the $\Lambda$'s only when all $\Phi$'s have been removed. As such, the decoder does fulfil the third requirement. It can therefore be applied to the case of Fibonacci anyons without alteration.

Even so, the fact that the decoder admits a finite threshold for the $\Phi-\Lambda$ anyons does not mean that it will also do so for the Fibonacci model. The non-Abelian anyons of these models have different behaviour, especially with regards to braiding. The effect of braiding for the $\Phi$ anyons merely represents their permutation, and has no further effect on the fusion space. The braiding of $\tau$ anyons, on the other hand, is universal for quantum computation. However, this difference can be expected to not have too much of an effect on the decoding.

To see why this is the case, consider the effect of the braiding non-Abelian anyons in general. If errors cause a pair of anyons to braid it will change the probabilities for fusion results when those anyons are fused with others. It does not change the probabilities for their fusion with each other. It is therefore equivalent to the pair not braiding, but instead having other pairs of anyons (for which the fusion product of this pair is the vacuum) coherently tunnelled between them.

The braiding of a pair of anyons that are separated by a distance $l$ will require $O(l)$ errors to occur. For an error model such as the one considered in this work, this means that the probability of such an braiding will decay as $(Cp)^l$ for some coefficient $C<1$. The probability that errors will tunnel anyons over the same distance decays similarly. So these two processes not only have similar effects but also occur with similar probability.

The fact that $\Phi$ anyons have trivial braiding should therefore not have a significant effect on their error-correctability, since they are subject to errors that have the same effect as non-trivial braiding and which occur with a similar probability. The fact that a finite threshold is found for the $\Phi-\Lambda$ in this work therefore strongly suggests the same for the Fibonacci anyons subject to a corresponding error model.



\begin{thebibliography}{99}
\bibitem{note} Note that the roles of vertices and plaquettes here are the opposite to other works. However, this is merely a convention and does not change the model.
\end{thebibliography}
\end{document}